\documentclass[12pt]{article}

\usepackage[colorlinks,citecolor=blue,linkcolor=blue]{hyperref}
\usepackage{graphicx}
\usepackage{booktabs} 
\usepackage[normalem]{ulem}

\usepackage{amsmath}
\usepackage{graphicx}
\usepackage{natbib, color}
\def\red{\color{black}}

\def\bzero{{\bf 0}}
\def\bone{{\bf 1}}
\def\br{{\bf r}}

\def\bu{{\bf u}}
\def\bv{{\bf v}}

\def\by{{\bf y}}

\def\bI{{\bf I}}

\def\bP{{\bf P}}
\def\bQ{{\bf Q}}

\def\bX{{\bf X}}
\def\bY{{\bf Y}}

\def\thick#1{\hbox{\rlap{$#1$}\kern0.25pt\rlap{$#1$}\kern0.25pt$#1$}}

\def\bbeta{\boldsymbol{\beta}}

\def\bepsilon{\boldsymbol{\epsilon}}

\def\smbalpha{\boldsymbol{{\scriptstyle{\alpha}}}}

\def\phat{{\widehat p}}

\def\bYhat{{\widehat \bY}}

\def\smbalpha{\widehat{\smbalpha}}

\def\hbar{\bar{ h}}

\def\Ybar{\bar{ Y}}

\def\transpose{{\sf \scriptscriptstyle{T}}}

\def\trans{^{\transpose}}

\def\mybox#1{\vskip1mm \begin{center}
        \hspace{.0\textwidth}\vbox{\hrule\hbox{\vrule\kern6pt
\parbox{.9\textwidth}{\kern6pt#1\vskip6pt}\kern6pt\vrule}\hrule}
        \end{center} \vskip-5mm}
\def\lboxit#1{\vbox{\hrule\hbox{\vrule\kern6pt
      \vbox{\kern6pt#1\vskip6pt}\kern6pt\vrule}\hrule}}

\def\thickboxit#1{\vbox{{\hrule height 1mm}\hbox{{\vrule width 1mm}\kern6pt
          \vbox{\kern6pt#1\kern6pt}\kern6pt{\vrule width 1mm}}
               {\hrule height 1mm}}}

\def\fat#1{\hbox{\rlap{$#1$}\kern0.25pt\rlap{$#1$}\kern0.25pt$#1$}}

\begin{document}

\def\spacingset#1{\renewcommand{\baselinestretch}%
{#1}\small\normalsize} \spacingset{1}

\title{ \bf Rediscovering a little known fact about the $t$-test 
    and the $F$-test:
  Algebraic, Geometric, Distributional and Graphical Considerations} 

\author{Jennifer A.\ Sinnott, Steven N.\ MacEachern, and Mario Peruggia \\ \small Department of Statistics, The Ohio State University, Columbus, Ohio, USA}

\maketitle

\begin{abstract}
\noindent We discuss the role that the null hypothesis should play in the
construction of a test statistic used to make a decision about that
hypothesis.  To construct the test statistic for a point null
hypothesis about a binomial proportion, a common recommendation is to
act as if the null hypothesis is true.  We argue that, on the surface,
the one-sample $t$-test of a point null hypothesis about a Gaussian
population mean does not appear to follow the recommendation.  We show
how simple algebraic manipulations of the usual t-statistic lead to an
equivalent test procedure consistent with the recommendation.   We
provide geometric intuition regarding this equivalence and we consider
extensions to testing nested hypotheses in Gaussian linear models.  We
discuss an application to graphical residual diagnostics where the
form of the test statistic makes a practical difference.  By examining
the formulation of the test statistic from multiple perspectives in
this familiar example, we provide simple, concrete illustrations of
some important issues that can guide the formulation of effective
solutions to more complex statistical problems. 
\end{abstract}


\newpage

\spacingset{1.45} 

\section{Introduction}
\label{intro}

Among the first procedures taught in an introductory statistics class
are hypothesis testing and confidence interval estimation for a
proportion (see, e.g., \cite{moore2012introduction}).  For example,
students may be 
given data on the sexes of a sample of $n$ babies born during a
certain time period.  They may be asked either to estimate the true
proportion $p$ of babies born male and provide a confidence interval,
or to test whether the proportion is equal to, for example,~0.5.\footnote{
There is evidence that this proportion is larger than 0.5 in
most of the world; (see, e.g., \cite{chao2019systematic}).}
Typically, for large $n,$ the distribution of the sample proportion is
approximated by
$\phat \stackrel{\cdot}{\sim} N\left(p, p(1-p)/n\right),$ and two
slightly different procedures are introduced.  For estimation and
confidence interval construction, $\phat$ is commonly plugged into the
variance formula, and a $100(1-\alpha)\%$ confidence interval is
calculated as
\begin{equation}
\phat \pm z_{\frac{\alpha}{2}} \sqrt{\phat(1-\phat)/n}.
\label{p_ci}
\end{equation}
For testing $H_0: p=p_0$ for a pre-specified $p_0$, students are advised to act as though the null were true, and use the null to {\em construct} the test statistic.  As a result, $p_0$ is
plugged into the variance formula, producing the test statistic \begin{equation}
  \frac{\phat - p_0}{\sqrt{p_0(1-p_0)/n}}.
\label{p_test}
\end{equation}
Although many different approaches to both testing and interval
estimation have been proposed --- {and many commonly used
  statistical software packages allow the user to apply continuity
  corrections to these formulas to improve the asymptotic
  approximation (e.g., by setting the argument {\tt
    correct = TRUE} in the R function {\tt prop.test})} --- in the
authors' experience, the above methods are still frequently taught for
hand calculation in introductory statistics classes of various levels.
For instance, Example~10.3.5 in \cite{casella2002statistical}
discusses precisely two test procedures based on test statistics that
use $\phat$ or $p_0$ to estimate the variance, commenting on their
relative merits in terms of a comparison of their power functions.
For further discussions of procedures used in the one-sample
proportion setting, see, e.g., \cite{agresti1998approximate} and
\cite{yang2019using}.

Also among the first procedures taught are estimation and hypothesis testing for the mean $\mu$ of a normal
$N(\mu, \sigma^2)$ population with unknown variance $\sigma^2.$ For
example, students may be given data on the heights of a random sample
of U.S.\ women and be asked to estimate the true mean height, or test
whether it is equal to some specified value.  If our data consist of a
random sample $Y_1, \ldots, Y_n$ from the $N(\mu, \sigma^2)$
population, $\Ybar \sim N\left(\mu, \sigma^2/n \right),$ and
a confidence
interval is constructed analogously to (\ref{p_ci}), as
$$\Ybar \pm t_{n-1, \frac{\alpha}{2}} S/\sqrt{n}$$
where 
\begin{equation}
S^2 = \frac{1}{n-1}\sum_{i=1}^n (Y_i - \Ybar)^2
\label{sample_var}
\end{equation}
is the sample variance.  (This follows from observing that 
$T  := (\Ybar - \mu)/(S/\sqrt{n})$ has a $t$ distribution
with $n-1$ degrees of freedom, accounting for the replacement of $\sigma$ with $S$.)
To test $H_0: \mu = \mu_0$ for a
pre-specified $\mu_0,$ we can, analogously to (\ref{p_test}), invoke the
null.  When $H_0$ holds, we know $\mu=\mu_0$ but still need to
estimate $\sigma^2.$ Since $\mu$ is known, the most efficient estimator of
$\sigma^2$ is:
$$S_0^2 := \frac{1}{n}\sum_{i=1}^n(Y_i - \mu_0)^2.$$
Our test statistic would thus be:
$$T_0  := \frac{\Ybar - \mu_0}{S_0/\sqrt{n}}.$$

But, of course, people do not use this test statistic!  Instead, they
construct a statistic that ignores the information that $\mu=\mu_0$
provided by $H_0,$ and perform the standard one-sample $t$-test using
the test statistic
$$T=\frac{\Ybar - \mu_0}{S/\sqrt{n}}.$$
At first glance, one might suspect that using this test statistic
would be less efficient than using $T_0,$ since its denominator has
$n-1$ degrees of freedom rather than $n.$

We are thus led to wonder
why information provided by the null is discarded in constructing the
one-sample $t$-test.  In the remainder of the paper we clarify this
question and present a more general perspective {that we think
  will be of interest to colleagues who teach this material as well as
  those interested in the development and implications of some of our
  most fundamental statistical tools.}

\section{Establishing the connection}
\label{sec:connection}
The connection between 
the two methods proposed at the end of
the previous section can be established from an algebraic and from a
geometric point of view.  We look at these two approaches separately. 

To begin, we note that any intuition that a test based on
$T_0$ rather than $T$ could be more efficient is wrong: a tail-area
test based on $T_0$ and one based on $T$ produce {\em identical}
answers.  This is because $T$ is a one-to-one, increasing function of
$T_0,$
\begin{equation}
T=\frac{\sqrt{n-1}~T_0}{\sqrt{n-T_0^2}}, \label{one_s_rel}
\end{equation}
over the interval $(-\sqrt{n},\sqrt{n})$, which is the set of possible
values for $T_0$.  Specifically, for any fixed $\alpha$, with
  $0 \leq \alpha \leq 1$, let $c_\alpha \geq 0$ be the critical value of the
  size $\alpha$ test based on $T_0$.  The rejection region of this test
  is $$R_{T_0} = \{\by = (y_1, \ldots, y_n)\trans : |T_0(\by)| \geq
  c_\alpha\}.$$  Because the transformation in
  Equation~\eqref{one_s_rel} is monotonic increasing on $[0,\sqrt{n})$, 
the set $$R_T = \{\by = (y_1, \ldots, y_n)\trans : |T(\by)| \geq
(\sqrt{n-1}~c_\alpha)/(\sqrt{n-c_\alpha^2})\}$$ satisfies $R_T = R_{T_0}$.  
It follows that the test that rejects if and only if 
$|T(\by)| \geq
(\sqrt{n-1}~c_\alpha)/(\sqrt{n-c_\alpha^2})$ has the exact same
rejection region (in sample space) as the test that rejects when 
$|T_0(\by)| \geq c_\alpha$.  The two tests must then have the same
size and 
power function and are therefore equivalent.

As noted by 
  a colleague, a
  simple way to establish Equation~\eqref{one_s_rel} is to recognize
  that the one sample $t$-test can be derived as a likelihood ratio
  test that rejects $H_0: \mu = \mu_0$ when the ratio
\[
  \lambda(\bY) =
  \frac{\sup_{\sigma^2} L(\mu_0,\sigma^2|\bY)}{\sup_{\mu,\sigma^2}
    L(\mu,\sigma^2|\bY)} 
\]
is small or, equivalently, when
the ratio of sums of squares under the null and full model,
\begin{equation}
R=\frac{\sum_{j=1}(Y_j-\mu_0)^2}{\sum_{j=1}^n(Y_j-\Ybar)^2},
\label{eq:LRT_ratio_t}
\end{equation}
is large.  This ratio can be expressed as
$$R=\frac{\sum_{j=1}(Y_j-\Ybar)^2 + n(\Ybar -
  \mu_0)^2}{\sum_{j=1}^n(Y_j-\Ybar)^2} = 1 + \frac{T^2}{n-1}$$
or as
$$R=\frac{\sum_{j=1}(Y_j-\mu_0)^2}{\sum_{j=1}^n(Y_j-\mu_0)^2 -
  n (\Ybar - \mu_0)^2} = \frac{1}{1-T_0^2/n}.$$ The former expression
leads to the standard $t$-test based on $T$, while the latter leads to
the test based on $T_0$.  Equating these two expressions yields the
identity of Equation~\eqref{one_s_rel}.  

This relationship between $T$ and $T_0$ is, of course, not new: for
example, it arises substantively in Lehmann's approach for
demonstrating that the one sample $t$-test is a uniformly most
powerful (UMP) unbiased test of $H_0: \mu=\mu_0$ vs.\
$H_A: \mu \neq \mu_0$ \cite{lehmann}.  The full details of the
argument are best left to Lehmann, but, very briefly, for parameters
in exponential family distributions, Lehmann's Theorem 1 in Chapter 5
gives a set of conditions about the form of a test statistic in
relation to the family's sufficient statistics.  When these conditions
are satisfied, a test based on the test statistic is UMP unbiased.
The set of conditions Lehmann provides is satisfied by $T_0$ rather
than $T,$ and the UMP unbiasedness of the $t$-test is then established
by exhibiting that $T$ is a one-to-one function of $T_0.$

Interestingly, this equivalence does not seem to be widely
  known (at least based on our informal surveying of several
  colleagues).  This is somewhat surprising. In fact, in addition to
  appearing in Lehmann's book, the algebraic equivalence of the test
  statistics is periodically mentioned in the literature (see,
  e.g.,
  \cite{lefante1986c257,good1986comments,shah1987c293,shah1993testing,lamotte1994note}). 
  However, we feel that the equivalence is worth revisiting, both in the context of the $t$-test and in the more general setting of nested linear models, where an analogous equivalence holds.  The geometric interpretation of the equivalence, not described in these earlier references, provides an interesting addition to the geometric interpretation of linear models. Moreover, despite the test statistics leading to identical conclusions in the linear models setting, one choice naturally leads a practitioner to consider so-called studentized residuals while the other leads to so-called standardized residuals---and these sets of residuals do have different properties and, when plotted, may lead to different visual interpretations.  We expand on these remarks in subsequent sections.
  
\section{The geometric point of view}
\begin{figure}
 \begin{minipage}[b]{0.5\linewidth}
\centering
\includegraphics[trim=200 560 200 100,clip,width=0.6\textwidth]{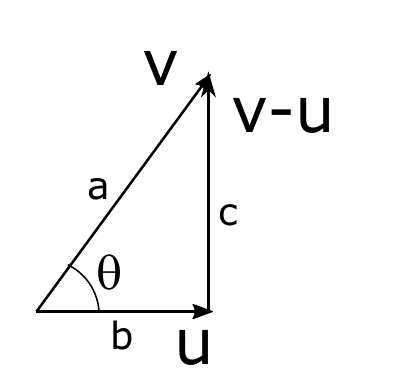}
\end{minipage}
\begin{minipage}[b]{0.5\linewidth}
$a = \Vert \bv \Vert = \sqrt{\text{SSTO}}$,\\
$b = a \cos \theta = \Vert \bu \Vert = \sqrt{\text{SST}}$,\\
$c = a \sin \theta = \Vert \bv - \bu \Vert = \sqrt{\text{SSE}}$,\\ 
$T_0^2 = n \, (b^2/a^2) = n \cos^2 \theta$,\\
$T^2 = (n - 1) \, (b^2/c^2) = (n - 1) \cot^2 \theta$.
\vspace{3mm}
\end{minipage}
\caption{Geometric representation of the test statistics $T_0$ and $T$.}
\label{fig:geometry}     
\end{figure}

Interestingly, the equivalence of $T_0$ and $T$
can be understood geometrically because they can both be viewed as
trigonometric functions of the same angle, and it is possible to
express any trigonometric function in terms of any other trigonometric
function, up to sign.  To see the geometric relationship, define the
vectors $\bv = (Y_1 - \mu_0, Y_2 - \mu_0, \ldots, Y_n - \mu_0)\trans$
and $\bone=(1, 1, \ldots, 1)\trans.$ Then, the orthogonal projection
of $\bv$ onto $\bone$ is $\bu = (\Ybar - \mu_0)\bone,$ and the
Pythagorean Theorem implies:
\begin{eqnarray*}
  \Vert \bv \Vert^2 &=& \Vert \bu \Vert^2 \hspace{.55in} +
                              \hspace{.2in} \Vert \bv  -
                              \bu \Vert^2,   \\ 
  \text{i.e., } \hspace{.2in} 
  \sum_{i=1}^n (Y_i - \mu_0)^2 &=& n (\Ybar
                                   -
                                   \mu_0)^2
                                   \hspace{.1in}
                                   +
                                   \hspace{.2in}
                                   \sum_{i=1}^n
                                   (Y_i -
                                   \Ybar)^2,
  \\ 
  \text{i.e., } \hspace{.75in} \text{SSTO} &=& 
                                              \text{SST} \hspace{.6in}
                                              +\hspace{.2in}  \text{SSE}, 
\end{eqnarray*}
where we introduce analysis of variance terminology, with SSTO,
SST, and SSE indicating the Sums of Squares for Total, Treatment, and
Error, respectively.  Thus, if we define $\theta$ to be the
angle between $\bone$ and $\bv,$ then: 
$$T_0^2 = n \, \frac{\text{SST}}{\text{SSTO}} =
n \cos^2 \theta \,\,\, \text{ and } \,\,\, T^2 = (n-1) \,
\frac{\text{SST}}{\text{SSE}} = (n-1) \cot^2 \theta.$$ {A stylized,
  two-dimensional representation of the essence of these geometric
  relationships is presented in Figure~\ref{fig:geometry}.}  Using
basic trigonometric expressions it is easy to derive the stated
algebraic relationship between $T$ and $T_0$.  In fact,
\[
  T^2 = (n - 1) \cot^2 \theta = (n-1) \, \frac{\cos^2 \theta}{\sin^2
    \theta} = (n-1) \, \frac{\cos^2 \theta}{1 - \cos^2
      \theta}.
\]
Substituting $\cos^2 \theta = T_0^2 / n$ into this expression and
taking square roots on both sides (making sure the signs match, as
they should) yields Equation~\eqref{one_s_rel}.

\section{Extension to linear models}
The results presented in the previous sections are not specific to the
$t$-test setting.  In fact, constructing a test statistic by invoking
the null hypothesis and constructing it in the ``traditional'' way
produces equivalent test procedures across a range of linear models.
This connection can be established by rewriting the two statistics as
functions of different terms in the orthogonal decomposition of the
sum of squares. 

\subsection{Nested models}\label{sec:nested}
For instance, consider the standard linear model
\[
\bY = \bX \bbeta + \bepsilon,
\]
where $\bY = (Y_1, \ldots, Y_n)\trans$ is a vector of observations,
$\bX_{n \times p}$ is a design matrix of rank $p < n$,
$\bbeta = (\beta_1, \ldots, \beta_p)\trans$ is a vector of regression
parameters, and $\bepsilon = (\epsilon_1, \ldots, \epsilon_n)\trans$
is an error vector with elements
$\epsilon_i \stackrel{\text{iid}}{\sim} N(0, \sigma^2)$.
{Suppose we wish to determine if a specific collection of $p_2$
  covariates in $\bX$ does not significantly contribute to the
  prediction of $\bY$ in the linear model.  We can formulate this
  question as a testing problem in which the null hypothesis states
  that the $p_2$ regression coefficients for these covariates are all
  zero.}  Without loss of generality we can assume that the parameters
of interest are the last $p_2 < p$ and rewrite the model as
\[
\bY = \bX_{1} \bbeta_1 + \bX_{2} \bbeta_2 +
\bepsilon,
\] 
where $\bX = [ \bX_1 \vert \bX_2 ]$ and
$\bbeta = (\bbeta_1\trans,\bbeta_2\trans)\trans$, with $\bbeta_i$ of
dimension $p_i$ for $i=1,2,$ and $p_1 + p_2 = p.$ The testing problem
concerning the nested model 
can then be stated as
$$H_0: \bbeta_2 = \bzero \,\,\, \text{ vs. } \,\, H_A: \bbeta_2 \neq
\bzero.$$ 

Both the ``traditional'' and the ``null hypothesis'' testing
procedures try to quantify the importance of the reduction in error
sums of squares that ensues from entertaining the full model rather
than the reduced model, but they differ in the comparison yardstick
they use.  The ``traditional'' procedure uses a yardstick based on the
full model.  The ``null hypothesis'' procedure uses a yardstick based
on the reduced model with $\bbeta_2 = \bzero$.

Geometrically, the statistics arise from a sequence of projections.
Specifically, define:
$$\bP_1 = \bX_1 (\bX_1\trans\bX_1)^{-1} \bX_1 \trans, \hspace{15mm}  \bQ_1 =
\bI - \bP_1,$$ 
and
$$\bP_{12} = \bX (\bX\trans\bX)^{-1} \bX \trans, \hspace{15mm} \bQ_{12} =
\bI - \bP_{12}.$$ The matrix $\bP_1$ operates an orthogonal projection
onto the space spanned by the columns of the reduced design matrix
$\bX_1$ and the matrix $\bP_{12}$ operates an orthogonal projection onto
the space spanned by the columns of the full design matrix $\bX$.
Under the reduced model, the vector of predicted values is
$$\bYhat_1 = \bP_1 \bY,$$ 
the vector of residuals is 
$$ \br_1 = \bY - \bYhat_1 = \bQ_1 \bY,$$
and the residual sum of squares is
$$ 
\text{SSE}_1 = \bY\trans \bQ_1 \trans \bQ_1 \bY = \bY\trans \bQ_1 \bY.$$ 
Similarly, under the full model, the vector of predicted values is 
$$\bYhat_{12} = \bP_{12} \bY,$$ 
the vector of residuals is 
$$ \br = \bY - \bYhat_{12} = \bQ_{12} \bY,$$ 
and the residual sum of squares is
$$ \text{SSE}_{12}= \bY\trans \bQ_{12} \bY.$$ 
The reduction in sums of squares ensuing from fitting the larger model
is given by
$$\text{SS}_{2 \mid 1} = \text{SSE}_1 - \text{SSE}_{12} = \bY\trans
(\bQ_1 - \bQ_{12}) \bY = \bY\trans (\bP_{12} - \bP_1) \bY.$$

The ``traditional'' procedure compares $\text{SS}_{2 \mid 1}$ to
$\text{SSE}_{12}$, the error sum of squares for the full model, while
the ``null hypothesis'' procedure compares $\text{SS}_{2 \mid 1}$ to
$\text{SSE}_1 = \text{SS}_{2 \mid 1} + \text{SSE}_{12}$, the error sum
of squares for the reduced model envisioned to hold under the null.
After adjusting for the degrees of freedom of the various sums of
squares, the resulting test statistics are
$$F_\text{trad} = 
\frac{\text{SS}_{2 \mid 1}/p_2}{\text{SSE}_{12}/(n-p)}
$$
and
$$
F_\text{null} =
\frac{\text{SS}_{2 \mid 1}/p_2}{\text{SSE}_{1}/(n - p_1)} = 
\frac{\text{SS}_{2 \mid 1}/p_2}{(\text{SS}_{2 \mid 1}
  + \text{SSE}_{12})/(n - p_1)},$$
respectively.

\subsection{Algebra, geometry, and distributional results}
\begin{figure}
\begin{minipage}[b]{0.4\linewidth}
\centering
\includegraphics[trim=200 560 200
100,clip,width=0.7\textwidth]{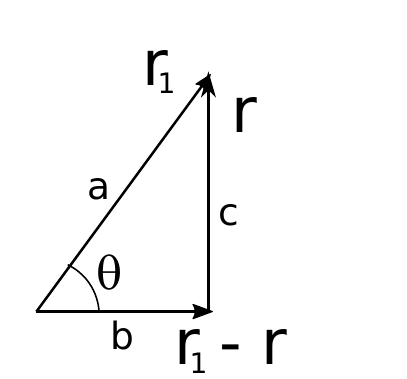} 
\end{minipage}
\begin{minipage}[b]{0.6\linewidth}
$a = \Vert \br_1 \Vert = \sqrt{\text{SSE}_1}$,\\
$b = a \cos \theta = \Vert \br_1 - \br \Vert = \sqrt{\text{SS}_{2 \mid 1}}$,\\
$c = a \sin \theta = \Vert \br \Vert = \sqrt{\text{SSE}_{12}}$,\\ 
$F_{\text{null}} = [(n-p_1)/p_2] \, (b^2/a^2) = [(n-p_1)/p_2] \cos^2 \theta$,\\
$F_{\text{trad}} = [(n-p)/p_2] \, (b^2/c^2) = [(n-p)/p_2] \cot^2 \theta$.
\vspace{3mm}
\end{minipage}
\caption{Geometric representation of the decomposition of the sums of
  squares for testing a nested hypothesis in the general linear
  model.}
\label{fig:geometry_F-test}       
\end{figure}

The orthogonal decomposition at play in this setting is analogous to
the one presented in Section~\ref{sec:connection} and is described
{in a stylized, two-dimensional display} in
Figure~\ref{fig:geometry_F-test}, along with the relationships between
its various elements.  Algebraic and trigonometric manipulations
similar to those outlined in
Section~\ref{sec:connection} show that $F_\text{trad}$ is a
one-to-one, increasing function of $F_\text{null}$ over
$(0,(n-p_1)/p_2)$, the set of possible values for $F_\text{null}$:
\begin{equation} \label{eq:F_stats_rel}
F_\text{trad} = \frac{(n-p) F_\text{null}}{n-p_1 - p_2 F_\text{null}}.
\end{equation}
Thus, as in the case of the $t$-test, tail-area tests using
$F_\text{trad}$ and $F_\text{null}$ are identical.
Note that, when $p=1,$ $p_1=0,$ and $p_2=1,$ the relationship between
$F_\text{trad}$ and $F_\text{null}$ given
in Equation~\eqref{eq:F_stats_rel}
reduces to the relationship between $T^2$ and $T_0^2$ implied by
Equation~\eqref{one_s_rel}.

The implementation of either test procedure requires knowledge of the
distribution of the corresponding test statistic under the null
hypothesis.  Using the notation introduced in
Figure~\ref{fig:geometry_F-test}, standard distributional results
imply that, under the null hypothesis, 
\begin{eqnarray*}
b^2 / \sigma^2&  = & \text{SS}_{2 \mid 1} / \sigma^2 \,\, \sim \,\,
                     \chi^2_{p_2},\\ 
c^2 / \sigma^2 & = & \text{SSE}_{12} / \sigma^2 \,\, \sim \,\,
                     \chi^2_{n-p}, 
\end{eqnarray*}
with $b^2$ independent of $c^2$.

Then, 
\[
F_\text{trad} = \frac{b^2/p_2}{c^2/(n-p)} \sim F_{p_2,n-p},
\]
as it is the ratio of two independent chi-square random variables
divided by their degrees of freedom. Also,
\[
\frac{p_2}{n-p_1} F_\text{null} =  \frac{b^2}{b^2 + c^2} \sim 
\text{Beta}\left(\frac{1}{2}\,p_2,\frac{1}{2}\,(n-p)\right), 
\]
as it is the ratio between a chi-square random variable and the sum
of that chi-square random variable and an independent chi-square
random variable.

\subsection{Does the difference ever matter?} \label{sec:residuals}
While the test procedures based on $F_\text{trad}$ and
  $F_\text{null}$ produce identical inferences, the realized values of
  the test statistics are different.  In this section we consider a
  situation in which, arguably, it is preferable to work with one of
  the two statistics rather than the other.

  Residual plots are effective graphical devices for assessing the
  quality of the fit of a linear regression model and for detecting
  potential outliers. As noted in Section~9.4.1 of
  \cite{weisberg2014applied}, a simple test for determining if
  observation $i$ is an outlier in a regression model that includes
  $p_1$ predictors is to include an additional predictor which is an
  indicator of the observation in question (i.e., a 0-1 vector whose
  only element equal to 1 is the $i$-th one) and to test if the
  regression coefficient of the indicator is equal to zero.

  Assuming normal errors for the regression model and letting
  $p_2 = 1$, it is natural to cast this problem into the framework of
  Section~\ref{sec:nested} and compare the full model with $p=p_1+p_2$
  predictors (the original predictors and the indicator of
  observation~$i$) and the nested model that omits the indicator
  variable.  Observation $i$ is declared an outlier if the null
  hypothesis that the coefficient of its indicator variable is zero is
  rejected.

  The traditional statistic for this problem is $F_\text{trad}$, which
  has an $F_{1,n-p}$ distribution under the null.  The square root of
  $F_\text{trad}$ (with sign matching the sign of the regression
  residual for observation $i$) is the usual $t$ statistic for outlier
  detection described by~\cite{weisberg2014applied}.  It is also a
  quantity known as the {\em studentized residual\/} for
  observation~$i$, a normalized version of the raw residual,
  {$\hat{e}_i$,} computed using an estimate of the error variance, 
  {$\hat{\sigma}^2_{(i)}$,} that {\em omits\/} observation~$i$
  from the calculation. Conceptually, this point of view is appealing
  because, if the null hypothesis were violated and observation $i$
  were indeed an outlier, its inclusion in the calculation would
  inflate the estimate of the error variance. {As stated
    in~\cite{weisberg2014applied}, the studentized residual can be
    expressed as
  \[
    t_i=\frac{\hat{e}_i}{\hat{\sigma}_{(i)}\sqrt{1-h_{ii}}},
  \]
  where $h_{ii}$ denotes the leverage of observation $i$ given by the
  $i$-th diagonal element of the {\red projection (or hat) matrix
    $\bP_{12}$ for the full model.}

On the other hand, as seen in Section~\ref{sec:nested}, the same test
could also be performed using the statistic $F_\text{null}$.  The
signed square root of $F_\text{null}$ turns out to be what is called
the {\em standardized residual\/} for observation~$i$, a normalized
version of the raw residual, {$\hat{e}_i$,} computed using an estimate
of the error variance, {$\hat{\sigma}^2$,} that uses all observations,
{\em including\/} observation~$i$.  This would be the natural
calculation to perform if one were to assume that the null hypothesis
were true.  {As stated in~\cite{weisberg2014applied}, the standardized
  residual can be expressed as
  \[
    r_i=\frac{\hat{e}_i}{\hat{\sigma} \sqrt{1-h_{ii}}},
\]
and the
  deterministic relationship between studentized and standardized
  residuals is given by
  \[
t_i = r_i \, \sqrt{\frac{n - p}{n - p + 1 - r_i^2}}
    \]
    This deterministic relationship mirrors, on the square root scale,
    the deterministic relationship between $F_\text{trad}$ and
    $F_\text{null}$.} Ultimately, because of the deterministic
  relationships relating $F_\text{trad}$, $F_\text{null}$, and the two
  residual test statistics, an outlier test based on any of these four
  statistics leads to the same decision.

  Residual plots are often used to conduct an exploratory assessment
  of the fit of the regression model.  In this type of analysis, the
  plots are scanned visually for the existence of identifiable
  patterns and idiosyncratic features that might reveal violations of
  the modeling assumptions.  With regard to outlier detection
  specifically, plots of residuals vs.\ fitted values are inspected to
  reveal the presence of unusually large residuals.  We argue that,
  owing to the nonlinearity of the transformation that relates
  standardized residuals to studentized residuals, a studentized
  residual plot is better suited than a standardized residual plot to
  achieve this goal.

  We illustrate this point with an example based on a subset of the
  data on brain and body weights for 100 species of placental mammals
  reported in~\cite{sacher1974relation}.  Here, for the measurements
  on the~21 species of primates included in the data set, we consider
  the simple linear regression of the natural logarithm of brain
  weight on the natural logarithm of body weight.  Standardized and
  studentized residual plots are presented in the top row of
  Figure~\ref{fig:mammals_residuals}.  Two species stand out: \textit{Homo
  Sapiens} (with large positive residuals) and \textit{Gorilla Gorilla} (with
  large negative residuals).  Both are flagged as outliers at the 0.05
  level with respective p-values of 0.0034 and 0.0301 (unadjusted for
  multiplicity of comparisons).

\begin{figure}
\centering
\vspace{-1.7cm}
\includegraphics[trim=0 0 0
0,clip,width=0.95\textwidth]{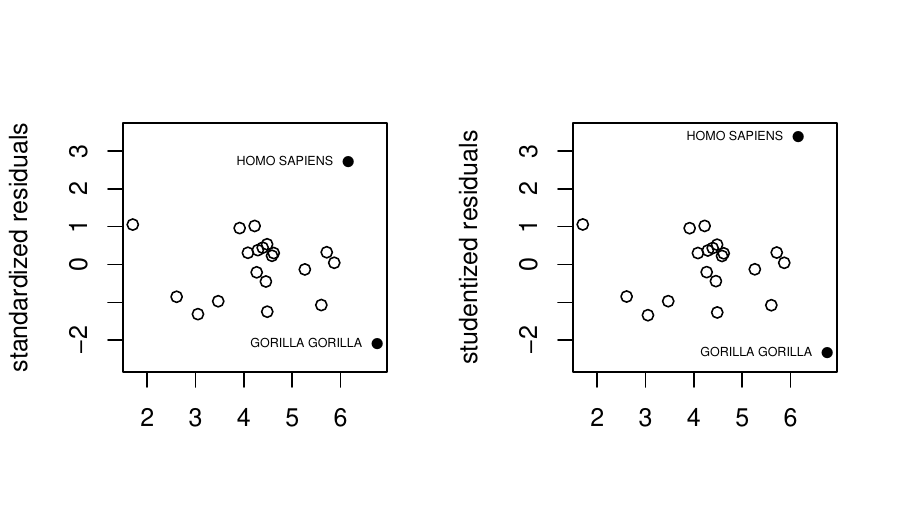} \\
\vspace{-2.5cm}
\includegraphics[trim=0 0 0
0,clip,width=0.95\textwidth]{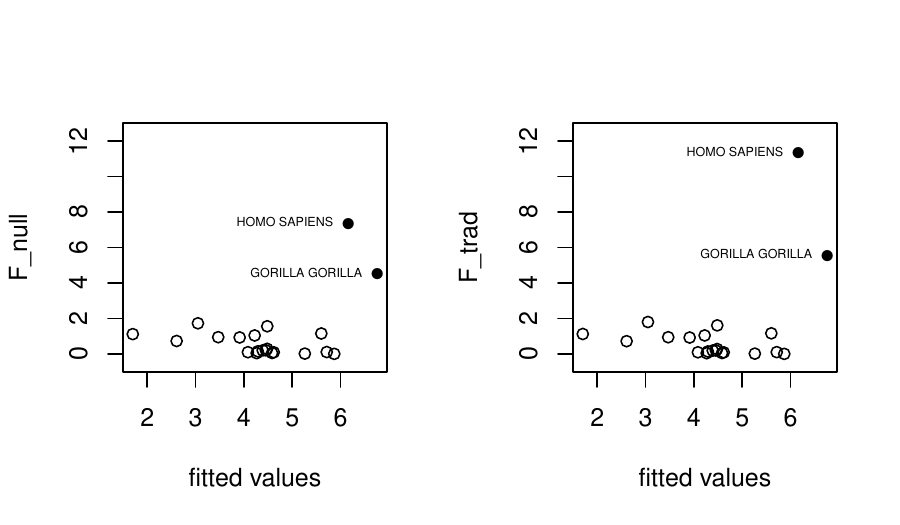} 
\vspace{-0.5cm}
\caption{Standardized and studentized residuals vs.\ fitted values for
  the primates data (top row) and their squared counterparts (bottom
  row).}
\label{fig:mammals_residuals}
\end{figure}

The extent to which these two species outlie compared to the other 19
species is clearly different.  As evidenced visually in both plots,
the residual for \textit{Homo Sapiens} is further removed from the bulk of the
residuals than the residual for \textit{Gorilla Gorilla} and this impression is
more notably accentuated in the studentized residual plot.  This is
due to the nonlinear relationship between standardized and studentized
residuals which causes the difference in absolute size between the two
to increase monotonically as the absolute size of the standardized
residual goes from 1 to infinity.  In particular,
{as shown in
  Figure~\ref{fig:mammals_residuals_diff},} 
the size of such difference becomes very noticeable
when the absolute value of the standardized residual exceeds a value
of about 2.5.

\begin{figure}
\centering
\vspace{-1.7cm}
\includegraphics[trim=0 0 0
0,clip,width=0.95\textwidth]{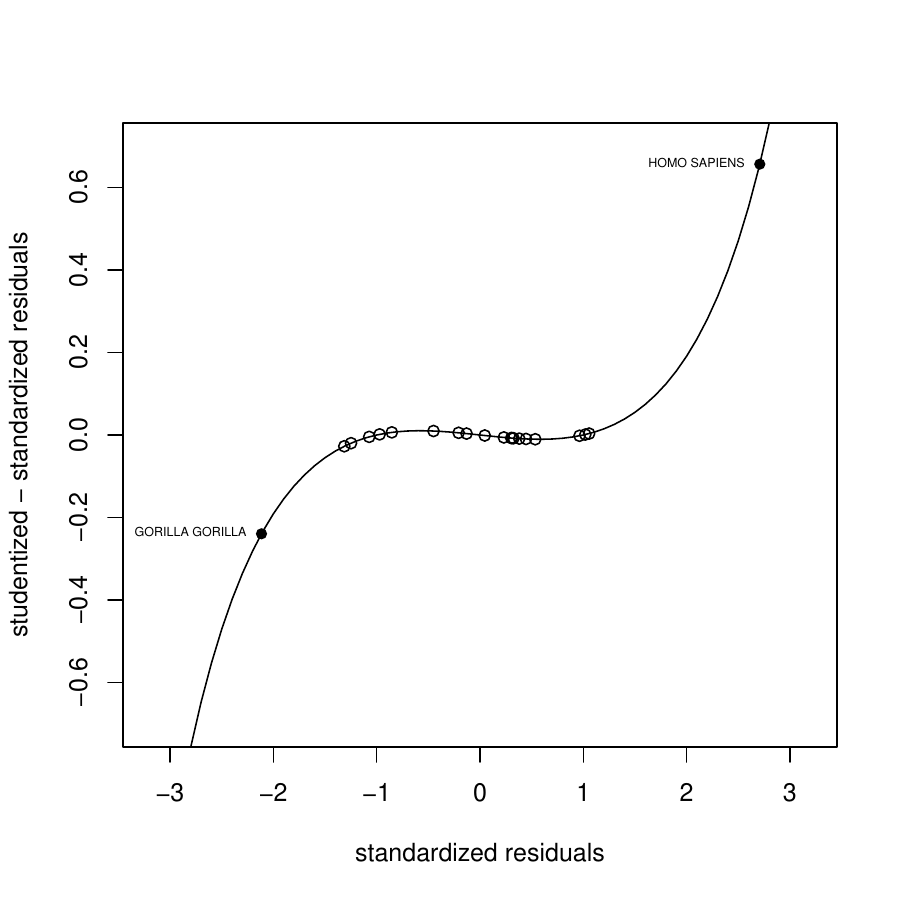} 
\vspace{-0.5cm}
\caption{{Differences between studentized and standardized residuals
  vs.\ standardized residuals for the primates data.  The solid line
  traces the deterministic relationship linking the plotted
  quantities.}  }
\label{fig:mammals_residuals_diff}
\end{figure}

In our example,
the absolute difference between studentized and standardized residuals
is 0.6563 (very noticeable) for \textit{Homo Sapiens}, 0.2394
(noticeable) for \textit{Gorilla Gorilla}, and between 0.0011 and
0.0273 (hardly noticeable) for all other species.  The displays in the
bottom line of Figure~\ref{fig:mammals_residuals}, being based on
$F_\text{null}$ and $F_\text{trad}$ which are the squared versions of
the standardized and studentized residuals, emphasize even more the
features just described.  In summary, the displays based on the
studentized residuals and on $F_\text{trad}$ can focus the analyst's
attention on the most extreme cases more effectively than those based
on the standardized residuals and on $F_\text{null}$.
  

\section{The Role of the Null Hypothesis in the Construction of a Test
  Statistic} \label{sec:role_of_null}

The fundamental question raised by the examples we presented {\red in}
this article concerns the role that the null hypothesis should play in
the testing paradigm.  By assumption, the null hypothesis is assumed
true in order to assess statistical significance, but to what extent
should one rely on it to {\em construct} the test statistic?  When
confronted with a new statistical model and a new parameter of
interest, it can be something of an art to determine a good choice of
test statistic.  Three common ``automatic'' approaches for
constructing test statistics from likelihoods privilege the null
differently: score tests are typically built under the null; Wald
tests are typically built under the alternative; and likelihood ratio
tests compare the null and the alternative somewhat equally.

{We consider first the case of an i.i.d.\ sample of size $n$ from
  $f(x \, | \, \theta)$, a {\red distribution} indexed by a single
  parameter, $\theta$, and rely on the results and examples presented
  in~\cite{casella2002statistical}. We denote by
  $L(\theta \, | \, \bX) = f(\bX \, | \, \theta)$ the likelihood
  function.

  The score is defined as
  $S(\bX \, | \, \theta) = d/d \, \theta \, \log f(\bX \, | \,
  \theta)$.  It can be shown that, for all $\theta$,
  $E S(\bX \, | \, \theta) = 0$ and
  $\text{Var} S(\bX \, | \, \theta) = I_n(\theta)$, the expected Fisher
  information in the sample.  The point null hypothesis
  $H_0: \theta = \theta_0$ is tested using the score test statistic
  $S(\bX \, | \, \theta_0)/\sqrt{I_n(\theta_0)}$, which has mean 0 and
  variance 1 for all $n$, and, under appropriate regularity
  conditions, converges in distribution under the null to a standard
  normal as $n$ 
  goes to infinity, enabling the derivation of approximate cut-off
  values.
  Equivalently, the test can be based on the square of the
  score test statistic which has an asymptotic $\chi^2_1$
  distribution.
{\red
  For $n$ independent Bernoulli$(p)$ observations yielding $y$
  successes, $\phat = y/n$ and 
  the resulting
  score test statistic for testing $H_0: p = p_0$
  is the one given in formula~(\ref{p_test}).
  Its squared version is therefore
  \[
   U_S(y,n;p_0) = \frac{(\phat - p_0)^2}{p_0 (1-p_0)/n}.
    \]
}
  
Suppose that, for all $\theta$, $W_n(\bX)$ is a consistent sequence of
estimators of $\theta$, having standard error $S_n(\bX)$.  The Wald
statistic for testing $H_0: \theta = \theta_0$ is constructed as
$(W_n(\bX) - \theta_0)/S_n(\bX)$ and, if asymptotic normality holds,
approximate cut-off values can again be derived under the null based
on the quantiles of a standard normal.  If the square of the Wald
statistic is used for testing, {\red approximate} cut-offs should
{\red be} based on the quantiles of a $\chi^2_1$ distribution.  Often
$W_n(\bX)$ is taken to be the maximum likelihood estimator of
$\theta$, with $S_n(\bX) = 1/\sqrt{I_n(W_n(\bX))}$.
{\red
Upon observing $y$ successes out of  
$n$ independent Bernoulli$(p)$ trials, this recipe
yields the
statistic of formula~(\ref{p_test}), but with $p_0$ replaced by
  $\hat{p} = y/n$
  in the denominator of that expression.  The squared
  version of the statistic is therefore
  \[
   U_W(y,n;p_0) = \frac{(\phat - p_0)^2}{\phat (1-\phat)/n}.
    \]
  }
  
  The likelihood ratio test statistic for testing $H_0: \theta =
  \theta_0$ is defined as 
\[
  \lambda(\bX) = \frac{L(\theta_0 \, | \, \bX)}{
    \sup_{\theta}L(\theta \, | \, \bX)}.
\]
Assuming appropriate regularity conditions, $-2 \log \lambda(\bX)$ has
an asymptotic $\chi^2_1$ distribution under the null that can be used
to obtain approximate cut-offs for the test.  For the case of $n$
independent Bernoulli$(p)$ observations, denoting by $y$ the total
number of successes, the resulting likelihood ratio test will reject
for {\red large} values of
\[
  {\red U_L(y,n;p_0) =}  -2 \log \left(\frac{p_0^y(1-p_0)^{n-y}}{
    {\hat{p}^y(1-\hat{p})^{n-y}}}
  \right).
  \]

  \cite{ENGLE1984775} defines these three types of tests for the more
  general situation in which the parameter vector is multidimensional,
  including the case in which only a subset of the parameters are of
  inferential interest while the remaining ones are regarded as
  nuisance parameters.  A detailed recount of the insightful results
  presented there is beyond the scope of this article, but an
  important message is that, quite generally, the three types of tests
  will behave asymptotically similarly under the null and under local
  alternatives, although the asymptotic behavior for alternative
  values away from $\theta_0$ will typically differ.

  For finite samples the three statistics may yield different
  tests. The reason for this is illustrated in
  Figure~\ref{fig:three_stats} which presents scatter plots of the
  squared score, {\red $U_S$}, and {\red squared} Wald, {\red $U_W$},
  statistics 
  against the log-likelihood 
  statistic, {\red $U_L$},  and of the squared score statistics, {\red
    $U_S$},  against the squared 
  Wald statistic, {\red $U_W$},  for $n=30$ and $p_0 = 1/3$. While
  these statistics 
  are, separately, related monotonically for $\hat{p} \le 1/3$ and
 $\hat{p} > 1/3$, the overall relationships are not monotonic. 
 An examination of the rejection regions for these tests shows
 {\red that} the order in which the total number of successes enters the 
 rejection region (as the size of the tests increase) differs among them.
 This is a situation in which the choice of which statistic to use matters.
  \begin{figure}
\centering
\vspace{-0.3cm}
\includegraphics[trim=0 0 0
0,clip,width=0.90\textwidth]{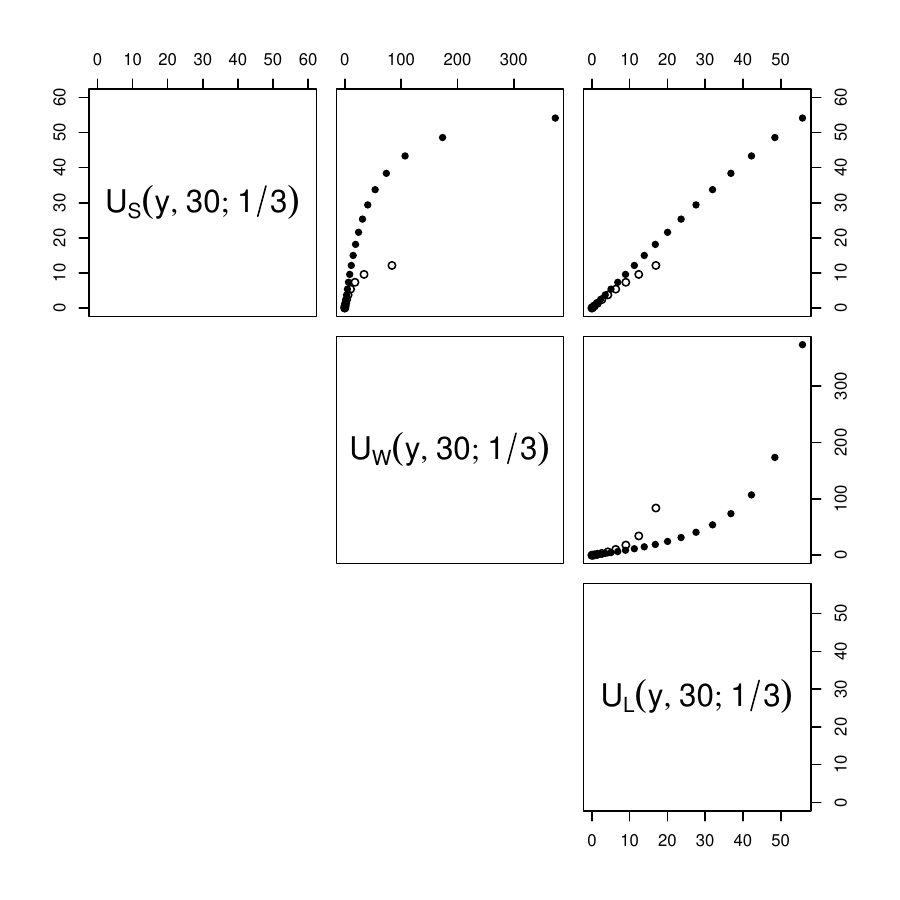} 
\vspace{-0.5cm}
\caption{Relationships between the {\red squared score, $U_S$,
    squared Wald, $U_W$,} and log-likelihood, {\red $U_L$,} test
  statistics for the case of independent Bernoulli data with $n=30$
  and $p_0=1/3$. {\red The open plotting symbols correspond to values
    of $y$ such that $\phat \leq 1/3$.  The solid plotting symbols
    correspond to values of $y$ such that $\phat > 1/3$.}  The
  statistics are not plotted for $y=0$ and $y=30$ to avoid cases where
  the Wald statistic is undefined and the log-likelihood statistic is
  close to minus infinity.}
\label{fig:three_stats}
\end{figure}

  As an example of a multidimensional situation including parameters
  of inferential interest and nuisance parameters, consider again the
  problem of testing a nested reduced model against the full model in
  the Gaussian linear model setting.  There, the likelihood ratio test
  rejects the null hypothesis that the reduced model holds when the
  ratio }
\[
  \lambda(\bY,\bX) = \frac{\sup_{\bbeta_1,\sigma^2}L(\bbeta_1,\sigma^2|\bY,
      \bX_1)}{\sup_{\bbeta,\sigma^2}L(\bbeta,\sigma^2|\bY, \bX)} 
\]
is small, or, equivalently, when the ratio
$\text{SSE}_{1}/ {\red \text{SSE}_{12}}$ of the error sum of squares under the
reduced (null) model and the full model is large, ultimately leading
to the equivalent tests based on $F_\text{null}$ {(a multiple of the
  score statistic as defined in \cite{ENGLE1984775})} and
$F_\text{trad}$ {(a multiple of the Wald statistic as defined in
  \cite{ENGLE1984775})}.  This structure of the likelihood ratio test
for nested models had already been noticed for the special case
presented in Section~\ref{sec:connection}, when discussing the
derivation of {the $t$-test in its two equivalent forms} based on the
ratio of Equation~\eqref{eq:LRT_ratio_t}.  {Using the multiparameter
  definitions of the three types of test statistics, their
  deterministic functional relationships, and considering their
  asymptotic and finite sample distributions, \cite{ENGLE1984775}
  shows that the resulting tests are, in this case, equivalent both
  asymptotically and in finite samples.}

\section{Discussion}
The idea of constructing a test statistic by pretending that the null
hypothesis is true is routinely presented as a general guideline when
using binomial data for testing the hypothesis that a population
proportion is equal to a given value.  Yet, this guideline is not
followed, at least on the surface, when normal data are used to build
the $t$-test for testing the hypothesis that the population mean is
equal to a given value.  As we noted in the paper, the $t$-test is
actually equivalent to a procedure based on a test statistic derived
by following the guideline, but making the connection requires a
little algebra, and is, to our knowledge, not typically made in
introductory statistics classes, even at the graduate level.  We have
also noted that the the same considerations presented for the $t$-test
extend to the use of the $F$-test for testing hypotheses concerning
nested linear models with Gaussian errors.

So, we are left to speculate why, in the case of the $t$-test and of
the $F$-test, the ``traditional'' procedure is preferred to the ``null
hypothesis'' procedure.  If a formal comparison is required, there is
no clear distributional advantage of one approach over the other.  For
the comparison of nested linear models, under the null, the
``traditional'' procedure requires calculation of the tail area of an
$F$ distribution and the ``null hypothesis'' procedure requires
calculation of the tail area of a Beta distribution.  If a power
calculation has to be performed under some alternative, it can be
based on the non-central $F$-distribution for the traditional
procedure and on the Type~I non-central Beta distribution for the
``null hypothesis'' procedure, again with no clear advantage of one
approach over the other.  Similar considerations apply to the case of
the $t$-test.

An appealing aspect of the ``traditional'' procedures is that the
$t$-statistic $T$ and the $F$-statistic $F_\text{trad}$ are both
constructed as ratios of independent quantities.  Because, in both
cases, the decision rule is based on an assessment of the relative
size of the numerator and denominator, it is conceivable that
independence may have been a key factor in establishing the tradition,
as an informal comparison of independent quantities is easier.  Under
the null, the denominators of the ``null hypothesis'' test statistics
are more efficient estimators of variability (have more degrees of
freedom) than their ``traditional'' counterparts.  However, this gain
in efficiency is offset by the dependence between numerator and
denominator (see \cite{lamotte1994note} for a related discussion).

In addition to the basic guiding principles, other considerations may
be at play when a certain tradition is established of preferring one
form of a test procedure over another for a given problem.  For the
nested model comparison, we already noted one desirable feature
exhibited by $F_\text{trad}$, namely that its numerator and
denominator are independent.  Another feature worth noting is that the
denominator of $F_\text{trad}$ does not depend on the particular
reduced model under consideration while the denominator of
$F_\text{null}$ does.  Although this is not much of a computational
burden, it is intuitively appealing to be able to use the same
yardstick in the denominator when testing different nested models
against the same full model.  Further, the graphical example of
  Section~\ref{sec:residuals} illustrates that when the value of the
  statistic itself is of interest, rather than the formal testing
  decision, there may be practical reasons for preferring the use of
  one statistic over the other.

  {In Section~\ref{sec:role_of_null} we reviewed three popular
    methods for building test statistics (the score, Wald, and
    likelihood ratio methods), discussing the different emphasis that
    they place on the null and alternative hypotheses.  For all cases
    examined in this paper, the three methods yield asymptotically
    equivalent procedures while emphasizing different features of the
    testing problem.  As noted in~\cite{ENGLE1984775} this is related
    to the different metrics used to evaluate discrepancy between the
    null and the alternative.  The Wald test accounts directly for
    differences in the parameter values, the likelihood ratio test
    measures differences in the log-likelihoods, and the score test
    assesses how steep the slope of the log-likelihood is at the null
    value.  While under very general conditions the three methods
    yield procedures that are asymptotically equivalent, we have
    noticed that the resulting finite sample tests may differ for
    independent Bernoulli data.  \cite{ENGLE1984775} presents
    additional examples where finite-sample conclusions might differ,
    comments on the different insight that the various formulations
    might bring to bear for specific models, and suggests that
    potential computational considerations might induce the analyst to
    opt for one of the tests over the other two.  }

  In sum, while we do not have a conclusive explanation as to why
  certain traditions have established themselves as the standard of
  practice for specific problems, we believe that these issues, often
  overlooked, are worth ruminating on, as they help us better see what
  considerations lead to the preference of one statistical procedure
  over another.  Choosing the right test statistic for a particular
  problem can be somewhat of an art, and understanding the
  similarities, differences, advantages, and disadvantages of the
  choice {in the simple settings we considered}
  may be helpful when turning to more
  complicated settings.

\section*{Acknowledgements} This material is based upon work supported
by the National Science Foundation under Grants No.~SES-1424481,
No.~DMS-1613110, and No.~SES-1921523.

\bibliographystyle{statistica} 
\bibliography{Bibliography-MM-MC}

\end{document}